\documentclass[preprint,showpacs]{revtex4}
\usepackage{graphicx} 
\begin{document}
\title{Universal and nonuniversal features in the crossover from linear to
nonlinear interface growth}
\author{T. J. Oliveira${}^{1,}$\footnote{Email address:
tiagojo@if.uff.br},
K. Dechoum${}^{1,}$\footnote{Email address:
kaled@if.uff.br},
J. A. Redinz${}^{2,}$\footnote{Email address:
redinz@ufv.br}
and F. D. A. Aar\~ao Reis${}^{1,}$\footnote{Email address:
reis@if.uff.br (corresponding author)}
}
\affiliation{
${}^{1}$ Instituto de F\'\i sica, Universidade Federal Fluminense,
Avenida Litor\^anea s/n, 24210-340 Niter\'oi RJ, Brazil \\
${}^{2}$ Departamento de F\'{\i}sica, Universidade Federal de Vi\c{c}osa,
36570-000 Vi\c{c}osa MG, Brazil}
\date{\today}

\begin{abstract}
We study a restricted solid-on-solid (RSOS) model involving deposition and
evaporation with probabilities $p$ and $1-p$, respectively, in one-dimensional
substrates. It presents a crossover from Edwards-Wilkinson (EW) to
Kardar-Parisi-Zhang (KPZ) scaling for $p\approx 0.5$. The associated KPZ
equation is analytically derived, exhibiting a coefficient $\lambda$ of the
nonlinear term proportional to $q\equiv p-1/2$, which is confirmed numerically
by calculation of tilt-dependent growth velocities for several values of $p$.
This linear $\lambda-q$ relation contrasts to the apparently universal
parabolic law obtained in competitive models mixing EW and KPZ components. The
regions where the interface roughness shows pure EW and KPZ scaling are
identified for $0.55\leq p\leq 0.8$, which provides numerical estimates of the
crossover times $t_c$. They scale as $t_c\sim \lambda^{-\phi}$ with $\phi =
4.1\pm 0.1$, which is in excellent agreement with the theoretically predicted
universal value $\phi =4$ and improves previous numerical estimates, which
suggested $\phi \approx 3$.
\end{abstract}
\pacs{PACS numbers: 05.40.-a, 05.50.+q, 68.55.-a, 81.15.Aa}
\maketitle

\section{Introduction}
\label{intro}

The competition between different
growth mechanisms is a characteristic of many real processes and has been the
subject of intensive investigation in the last years.
Many authors considered competitive growth models in which different dynamic
rules are randomly chosen for the aggregation of the incident particles
\cite{albano1,albano2,shapir,julien,tales,chamereis,muraca,lam,rdcor,
kolakowska,bulavin} and applications to real systems were suggested
\cite{shapir,julien,yu}. Such simplified models may mimic, for instance, the
effects of large energy distributions of the incident atoms, which lead to
different dynamic behavior as they arrive at the film surface.
They usually show crossover effects from one dynamics at small
times $t$ or short length scales $L$ to another dynamics at long $t$ or
large $L$.

In many cases, a crossover from the Edwards-Wilkinson (EW) \cite{ew}
dynamics to Kardar-Parisi-Zhang (KPZ) growth \cite{kpz} is observed. The
Langevin-type equation
\begin{equation}
{{\partial h}\over{\partial t}} = \nu{\nabla}^2 h + {\lambda\over 2}
{\left( \nabla h\right) }^2 + \eta (\vec{x},t) ,
\label{kpz}
\end{equation}
known as KPZ equation, is a hydrodynamic description of kinetic
surface roughening, where $h$ is the height at the position $\vec{x}$ in a
$d$-dimensional substrate at time $t$, $\nu$ represents a surface tension,
$\lambda$ represents the excess velocity and $\eta$ is a Gaussian
noise~\cite{kpz,barabasi} with zero mean and variance $\langle
\eta\left(\vec{x},t\right) \eta\left(\vec{x'},t'\right) \rangle = D\delta^d
\left(\vec{x}-\vec{x'}\right) \delta\left( t-t'\right)$. When $\lambda = 0$ in
Eq. (\ref{kpz}), we obtain the (linear) EW equation. Thus,
if $\lambda$ is very small, the features of EW growth are expected at small
times, and a crossover to KPZ behavior is observed at a characteristic time
$t_c$, when the macroscopic properties are affected by the overall nonlinear
character of the process. In this paper, we will analyze universal and
nonuniversal features of this crossover in lattice models through analytical
and numerical methods.

The roughness (or interface width) $W(L,t)$ is the simplest quantity that
indicates crossover effects. In lattice models, it is defined as
\begin{equation}
W(L,t) = \left< { \left[
{ {1\over{L^d}} \sum_i{ {\left( h_i - \overline{h}\right) }^2 } }
\right] }^{1/2} \right>
\label{defw}
\end{equation}
for deposition in a $d$-dimensional substrate of length $L$ ($h_i$ is
the height of column $i$ at time $t$, the bar in $\overline{h}$ denotes a
spatial average and the angular brackets denote a configurational average).
In a typical EW or KPZ system, it scales for small times as 
\begin{equation}
W\sim t^{\beta} .
\label{defbeta}
\end{equation}
However, when the crossover EW-KPZ is present, the
roughness exhibits two growth regions, characteristic of EW and KPZ
scaling ($\beta_{EW}<\beta_{KPZ}$ in any dimension), as shown qualitatively in
Fig. 1. At long times, the roughness saturates as
\begin{equation}
W_{sat}\sim L^{\alpha} .
\label{defalpha}
\end{equation}
$t_\times$ is the crossover time to the steady state or saturation regime, also
shown in Fig. 1.

From plausible scaling arguments (reviewed in Sec. III), several authors
suggested that, in $d=1$, the EW-KPZ crossover takes place for small $\lambda$
at
\begin{equation}
t_c \sim \lambda^{-\phi} ,
\label{deffi}
\end{equation}
with a universal crossover exponent $\phi =4$ \cite{AF,GGG,NT,forrest}.
However, to our knowledge the best known numerical estimate of this crossover
exponent is $\phi\approx 3$. It was
obtained by Guo, Grossman and Grant \cite{GGG} and by Forrest and Toral
\cite{forrest} through
numerical solutions of the KPZ equation and data collapse methods. Recent works
on lattice models confirmed the expected scaling relations for the growth and
saturation regimes of KPZ, even in the presence of the EW-KPZ crossover
\cite{chamereis}, but they were not able to improve the results for the EW
regime ($t\ll t_c$) or the crossover regions ($t\sim t_c$). Thus, neither a
numerical confirmation nor a thoroughly justified refutation of the
universality of the exponent $\phi=4$ was presented yet.

On the other hand, an universal relation between the coefficient $\lambda$ and
parameters of competitive lattice models with the EW-KPZ crossover was recently
proposed by  Braunstein and co-workers \cite{muraca,lam}. They considered
processes where the aggregation of incident particles followed the rules of a
KPZ lattice model with probability $p$ and the rules of an EW model with 
probability $1-p$. The most studied representative \cite{chamereis,julien} is
the competitive model involving ballistic deposition (BD - KPZ class)
\cite{vold} and the Family model, also known as random
deposition with surface relaxation (RDSR - EW class) \cite{family}. The
derivation of the corresponding KPZ equation from the stochastic rules of this
class of models gives $\lambda\sim p^2$ for small $p$ and is confirmed
numerically for the RDSR-BD model \cite{chamereis}.

In the present paper, we will study analytically and numerically a lattice model
with the crossover EW-KPZ in $d=1$, which is helpful to clarify the
universal and nonuniversal relations in this crossover. 
The model is a restricted solid-on-solid (RSOS)
one \cite{kk}, in which deposition and evaporation of particles compete with
probatilities $p$ and $1-p$, respectively. EW behavior is found for $p=1/2$, and
KPZ behavior for $p\neq 1/2$. We will derive analitically the KPZ equation for
this process, which exhibits $\lambda\sim q\equiv p-1/2$, where $q$ represents a
small relative probability of KPZ growth in the crossover region ($p\approx
1/2$, $q\approx 0$). This linear relation between $q$ and $\lambda$ is
confirmed numerically, and contrasts to the parabolic law found in other
competitive models. Consequently, the $\lambda-p$ relation in the EW-KPZ
crossover is clearly a model-dependent feature and not an universal law. On the
other hand, our numerical work will also provide an estimate of the crossover
exponent $\phi$ which agrees with the theoretically predicted universal value
$\phi=4$, improving previous estimates which failed to confirm that prediction.
This exponent is obtained from the scaling of $t_c(q)$, which is estimated from
the intersection of the EW and KPZ behaviors, as illustrated in Fig. 1.
The inherent difficulties of the numerical work, combined with the relatively
simple, linear $\lambda-p$ relation, explain why estimating the crossover
exponent is usually so hard.

The rest of this work is organized as follows.
In Sec. II, we will define precisely the discrete model, 
analitically derive its associated KPZ equation and confirm numerically the
$\lambda\sim q$ relation. In Sec. III we will review the
scaling arguments predicting $\phi =4$ and show the details of the numerical
analysis which gives $\phi=4.1\pm 0.1$. In Sec. IV we summarize our
results and present our conclusions. 

\section{The discrete model and the associated KPZ equation}
\label{secconnection}

In our competitive model, the deposit obeys the RSOS condition at any time, i.e.
the maximum height difference between neighboring columns is equal to the
particle size $a$ \cite{kk}. In the simulations, we consider $a=1$. At each
step of the process, a column of the deposit is randomly chosen. Subsequently,
deposition and evaporation attempts are chosen with probabilities $p$ and
$1-p$, respectively. When evaporation is chosen, the top particle of that
column is removed if the RSOS condition is satisfied after evaporation,
otherwise this attempt is rejected. When deposition is chosen, a new particle is
deposited at the top of that column if the RSOS condition is satisfied,
otherwise this attempt is rejected. The time unit $\tau$ corresponds to $L$
attempts of evaporation or deposition in a substrate with $L$ columns. In the
simulations, we considered $\tau=1$. 

This model was previously studied numerically by Amar and Family \cite{AF2}, in
order to show the universality of scaling functions and amplitude ratios for
KPZ processes in $d=1$. However, that analysis was restricted to $p\geq 0.75$,
consequently far from the region of EW-KPZ crossover.

Now we construct the associated KPZ equation of this process starting from the
master equation and performing a Kramers-Moyal expansion \cite{vankampen}, 
following the standard method used in
Refs. \cite{vvedensky,vvedensky1,muraca}.

First consider the deposition process according to the RSOS condition.
The transition rate $W(H,H')$ from the height configuration $H\equiv
\{ h_i\}$ to the configuration $H'\equiv \{ h_i'\}$ for this process is
\begin{equation}
W(H,H') = {1\over \tau} \sum_k{ w_k^{(0)} \delta \left( h_k', h_k+a \right)
\prod_{j\neq k}{ \delta \left( h_j' , h_j\right) } } ,
\label{probtrans}
\end{equation}
where the $\delta$ functions represent the condition that only the height of the
column of incidence can be increased and $w_k^{(0)}$ describes the condition
for aggregation:
\begin{equation}
w_k^{(0)} = \Theta\left( h_{k+1}-h_k \right) \Theta\left( h_{k-1}-h_k \right) ,
\label{wk0}
\end{equation}
where the $\Theta(x)$ is the unit step function, defined as $\Theta(x)=1$ for
$x\geq 0$ and $\Theta(x)=0$ for $x<0$.
Consequently, the first and the second transition moments are
\begin{equation}
K_i^{(1)} = \sum_{H'}{\left( h_i'-h_i\right) W\left( H,H'\right) }
= {a\over\tau} \Theta\left( h_{i+1}-h_i \right) \Theta\left( h_{i-1}-h_i
\right)
\label{mom1}
\end{equation}
and
\begin{equation}
K_{ij}^{(2)} = \sum_{H'}{\left( h_i'-h_i\right) \left( h_j'-h_j\right)
W\left( H,H'\right) }
= {a^2\over\tau} \Theta\left( h_{i+1}-h_i \right) \Theta\left( h_{i-1}-h_i
\right) \delta{\left( i,j\right) } .
\label{mom2}
\end{equation}

For RSOS evaporation, the transition rate and the transition moments are those
of Eqs. (\ref{wk0}), (\ref{mom1}) and (\ref{mom2}) with opposite signs in the
arguments of the $\Theta$ functions.

The Kramers-Moyal expansion of the master equation for the process provides
the stochastic equation \cite{vankampen}
\begin{equation}
{\partial{h_i}\over\partial{t}} = K_i^{(1)} + \sum_j{ \sqrt{K_{ij}^{(2)}}
\eta_j } ,
\label{langevingeral}
\end{equation}
where $\eta_j$ is a Gaussian white noise with zero mean and co-variance $\langle
\eta_i(t)\eta_j(t')\rangle = \delta {\left( i,j\right) }\delta(t-t')$.
For the competitive model, we obtain
\begin{equation}
{\partial{h_i}\over\partial{t}} = p {a\over\tau} \Theta\left( h_{i+1}-h_i
\right) \Theta\left( h_{i-1}-h_i \right) - (1-p) {a\over\tau}
\Theta\left( h_i-h_{i+1} \right) \Theta\left( h_i-h_{i-1} \right) + D_i\eta_i ,
\label{langevincomp}
\end{equation}
where $D_i$ is constant.

In order to pass from the discrete description of the model to its continuum
limit, we can use some analytical 
representation of the step function, which works in some limits. Many
regularization for the theta step function 
have already been suggested, such as the hyperbolic tangent function
\cite{park}, and maximum function \cite{vvedensky}. This representation is
expanded in Taylor series, so that
\begin{equation}
\theta (x) = c_{0} + c_{1}\;x + c_{2}\;x^{2} + \dots
\label{exptheta}
\end{equation}
Inserting this expansion in equation (\ref{langevincomp}), we get
\begin{eqnarray}
\frac{dh_i}{dt} &=& p \frac{a}{\tau} \left[ c_{0} +  c_{1} (h_{i+1} - h_{i}) + 
c_{2}(h_{i+1} - h_{i})^{2} \right] 
\left[ c_{0} +  c_{1} (h_{i-1} - h_{i}) +  c_{2}(h_{i-1} - h_{i})^{2} \right]
\nonumber \\
&-& (1-p)\frac{a}{\tau}
\left[ c_{0} +  c_{1} (h_{i} - h_{i+1}) +  c_{2}(h_{i} - h_{i+1})^{2} \right] 
\left[ c_{0} +  c_{1} (h_{i} - h_{i-1}) +  c_{2}(h_{i} - h_{i-1})^{2} \right]
\nonumber \\
&+& D_{i} \eta_{i}
\label{hi}
\end{eqnarray}

In the continuum limit, $a\to 0$. In this limit, $ac_1$ tends to a
finite, nonzero value, since the angular
coeficient $c_{1}$ in the expansion of the theta function  (Eq.
\ref{exptheta}) is of order $1/a$. Moreover, 
$a/\tau\to const$, since this is the random growth velocity.
We replace $h_{i}(t)$ by a smooth function
$h(x,t)$, whose coarse-grained derivatives are
\begin{eqnarray}
h_{i+1} - 2 h_{i} + h_{i-1} &\simeq& a^{2} \nabla^{2} h(x) \nonumber \\
h_{i+1} -  h_{i} &\simeq& a \nabla h(x)
\end{eqnarray}
Substitution in Eq. (\ref{hi}) gives
\begin{eqnarray}
\frac{dh}{dt} &=& (2p-1) \frac{a}{\tau} c_{0}^{2} + \frac{a^{3}}{\tau} c_{0}
c_{1}  \nabla^{2} h(x) + \nonumber \\
&+& (1-2p) \frac{a^{3}}{\tau} \left( c_{1}^{2} - 2 c_{0} c_{2} \right )  |
\nabla h(x) |^{2} + O(a^{5}) + \eta (x,t)
\label{ht}
\end{eqnarray}

This equation must reproduce correctly the random deposition model, 
when all interactions between columns are turned off. Consequently, the best
choice is to put $c_{0}=1$. Also, all the above mentioned choices for
representing the theta function, such as the hyperbolic tangent, lead to
$c_{2}=0$. This is typical of odd functions, such as $f(x)\equiv\theta (x) -
1/2$. Thus we get
\begin{equation}
{\partial{h}\over\partial{t}}(\vec{x},t) = \left( 2p-1\right) {a\over\tau} +
c_1 {a^3\over\tau} {\nabla^2 h} + \left( 1-2p\right) {c_1}^2 {a^3\over\tau}
{\left( \nabla h\right)}^2 + \eta (\vec{x},t) ,
\label{kpzrsoscomp}
\end{equation}
which is the KPZ equation associated with the RSOS model
with deposition and evaporation.
All terms in the right hand side of Eq. (\ref {kpzrsoscomp})
are finite quantities because $a/\tau$, $a c_{1}$, $\nabla h$ and 
$a\nabla^{2} h$ are expected to have the same order of magnitude.

It is interesting to recall that the choice of the value of $\theta(0)$ is
arbitrary because the step function is nonanalytic at the
origin. Thus, if our choice were $\theta(0)=1/2$, instead Eq. (\ref{wk0}) we
would have to represent the aggregation condition as
\begin{equation}
w_{k}^{(0)}= \Theta (h_{k+1}-h_{k}) \Theta (h_{k-1}-h_{k})
\left[ 1 + \delta{\left( h_k,h_k+1\right) } + \delta {\left( h_k,h_k-1\right) }
+ \delta {\left( h_k,h_k+1\right) }\,\delta {\left( h_k,h_k-1\right) }
\right ] ,
\end{equation}
where $\delta {\left( i,j\right)}$ is the discrete Kronecker delta.
As expected, this also gives the KPZ equation for the model, but
the choice $\theta(0)=1$ is suitable to represent the aggregation rule in a
concise form.

Comparison of Eqs. (\ref{kpzrsoscomp}) and (\ref{kpz}) shows that $\lambda$
varies linearly with $q\equiv p-1/2$. As expected, $\lambda <0$ when deposition
is dominant, and $\lambda >0$ for dominant evaporation. 
Such linear relation is similar to that predicted for a single step model by
Derrida and Mallick \cite{derrida} through a mapping into a one-dimensional
asymmetric exclusion model. On the other hand, it contrasts to the $\lambda\sim
p^2$ law obtained by Muraca et al \cite{muraca} for pure KPZ models (with
finite $\lambda$) competing with pure EW models, such as the RDSR-BD model
\cite{chamereis}.

This linear $\lambda-p$ relation was confirmed numerically.
The coefficient $\lambda$ can be calculated from the tilt-dependent
growth velocity in the KPZ regime. If a given KPZ process takes place on an
infinitely large substrate of inclination $u$, then $\lambda$ is related to the
growth velocity $v$ as \cite{krugmeakin,krugspohn,huse}
\begin{equation}
\lambda = {\left( {{\partial^2 v}\over{\partial u^2}}\right)}_{u=0} 
\label{calclambda}
\end{equation}
(this form applies to $d=1$, but is straightforwardly extended to higher
dimensions).
Several probabilities in the range $0.55\leq p\leq 1$ were considered for the
simulations in substrates of length $L={10}^4$. For each $p$, inclinations
from $u=0.1$ to $u=0.8$ were considered, and the deposit was
grown until times sufficient long for the KPZ regime to be attained. Average
values were taken over $100$ realizations for each $p$ and $u$.

Fig. 2a illustrates the method to calculate $\lambda$ from the growth velocities
for three different values of $p$. The parabolic fits accurately represent the
data behavior for all inclinations. Using those fits and Eq.
(\ref{calclambda}), we obtained estimates of $\lambda$ for each $p$. In order to
check the accuracy of these estimates, we also calculated the ratio
$\left( v-v_0\right) /u^2$ ($v_0$ is the growth velocity at zero slope), and
extrapolated that ratio to the limit $u\to 0$. The estimates of $\lambda$
agreed with those obtained from the parabolic fits within error bars.
In Fig. 2b we show $\lambda$ versus $q$, which confirms the linear relation
between those quantities for a large range of values of $q$, in agreement with
the KPZ equation obtained for the process.

\section{Numerical study of the EW-KPZ crossover}
\label{seccrossover}

First we recall the arguments that lead to the prediction of a crossover
exponent $\phi=4$.

In the works of Grossmmann, Guo and Grant (GGG) \cite{GGG} and of Nattermann and
Tang (NT) \cite{NT} (see also \protect\cite{forrest}), this result is derived
from multiscaling relations for systems with crossover EW-KPZ in $d=1$.
They proposed relations in the form
\begin{equation}
W(L,t) = L^{\alpha} f \left( {\frac{t}{t_c}, \frac{L}{\xi_c} } \right) ,
\label{multiscaling}
\end{equation}
where $\xi_c \sim t_c^{1/z_{EW}}$ is a crossover length and $z_{EW}=2$ is the
dynamical exponent of EW processes.
Assuming that $t_c$ scales as Eq. (\ref{deffi}), they obtained $\phi = z_{EW} /
(\alpha_{EW} + z_{EW} - 2 )$ using scaling arguments. In $d=1$, we have
$\beta_{EW}=1/4$ and $\alpha_{EW}=1/2$, which gives $\phi= 4$ in $d=1$.
This was confirmed by one-loop renormalization group calculations by NT.

The same result also follows from the expected roughness scaling of KPZ in
$d=1$. Assuming dynamic scaling in the nonlinear and saturation regimes, Amar
and Family \cite{AF,AF2} showed that the roughness scales as
\begin{equation}
W(L,t) \sim L^{1/2}  g\left( |\lambda| \frac{t}{L^{3/2}}  \right) ,
\label{fvkpz}
\end{equation}
where $g$ is a scaling function and where the dependence of $W$ on the
parameters $\nu$ and $D$ of Eq. (\ref{kpz}) was omitted. In the growth regime,
it gives
$W\sim {|\lambda|}^{1/3} t^{1/3}$ ($\beta_{KPZ}=1/3$ in Eq. \ref{defbeta}). Now
assuming that the crossover EW-KPZ takes
place when the EW roughness (Eq. \ref{defbeta} with $\beta_{EW}=1/4$) matches
that of KPZ, as illustrated in Fig. 1, we obtain $\lambda^{1/3} {t_c}^{1/3}\sim
{t_c}^{1/4}$, from which $\phi=4$ also follows.

This last argument is the basis for the numerical calculation of $t_c$ using the
roughness in the EW and KPZ regimes.
First, it is necessary to calculate scaling amplitudes not shown in Eq.
(\ref{defbeta}): for the EW regime we have
\begin{equation}
W_E \approx At^{1/4} ,
\label{we}
\end{equation}
and for the KPZ regime we have
\begin{equation}
W_K \approx Bt^{1/3} .
\label{wk}
\end{equation}
Matching these forms at $t_c$ we obtain
\begin{equation}
t_c \approx {\left( {A\over B}\right) }^{12} .
\label{tcab}
\end{equation}
Consequently, $t_c$ can be determined from the estimates of the amplitudes $A$
and $B$.

Simulations of the RSOS model with deposition and evaporation were done for
several values of $p$, in lattices with $L={10}^5$, up to times approximately
${10}^6$. $100$ deposits were generated for each $p$.

In Fig. 3 we show $W/t^{1/4}$ for small times $t$ with $p=0.7$ and $p=0.6$. That
ratio is
expected to be constant in the EW regime. A narrow region  $20\leq
t\leq 40$ with this feature is observed for $p=0.7$, while a wider EW region is
found for smaller $p$. Here it is important to recall
that other competitive models fail at this point because a clear EW region is
found only for very small $p$, where the KPZ regime becomes difficult to be
attained in simulation; one example is the model involving ballistic deposition
and the Family model studied in Ref. \protect\cite{chamereis}.

The calculation of amplitude $B$ is slightly more difficult because the ratios
$W/t^{1/3}$ are not constant inside a time window long enough to extend to the
maximum simulation times. In other words, the presence of significant
corrections to scaling in Eq. (\ref{wk}) has to be taken into account. This can
be done with the extrapolation of $W/t^{1/3}$ as a function of $1/t^{1/3}$, as
shown in Fig. 4 for $p=0.7$ and $p=0.6$ (see also Ref.
\protect\cite{chamereis}).
Although the range of the variable $1/t^{1/3}$
(abscissa of Fig. 4) is relatively small, it comprises almost two decades of
the largest values of $t$. Good linear fits of the data are obtained in these
large
time regions, which suggests constant (but large) subdominant corrections to
Eq.(\ref{wk}). The amplitude $B$ is estimated from the intersection of those
fits with the vertical axis ($t\to\infty$).

For fixed $p$, different ranges of the variable $1/t^{1/3}$ were chosen for the
extrapolation of the data and the calculation of error bars in the estimates of
the amplitude $B$. This procedure provides
reliable and accurate final estimates of that amplitude. For instance, for
$p=0.7$ (Fig. 4), we obtain $B=0.330\pm 0.004$.
We also observe that, while the amplitude $A$ slowly varies with $p$ (nearly
$10\%$ from $p=0.55$ to $p=0.8$), the amplitude $B$ has a remarkable dependence
on $p$.

The estimates of $t_c$ obtained from Eq. (\ref{tcab}) are shown in Fig. 5 as a
function of $q\equiv p-1/2$. Linear fits of different subsets of those data give
\begin{equation}
t_c\sim q^{-\left( 4.1\pm 0.1 \right)} .
\label{tcq}
\end{equation}
The linear relation between $q$ and $\lambda$ implies $\phi=4.1\pm 0.1$, which
is in excellent agreement with the theoretically predicted value.

Here it is important to recall that, in other competitive models such as the
RDSR-BD one \cite{chamereis}, the calculation of $t_c$ with accuracy was not
possible. For instance, a clear EW growth regime (with $\beta_{EW}=1/4$) is
observed in that model only for very small $p$, but in these conditions the KPZ
growth regime (with $\beta_{KPZ}=1/3$) is not attained within a reasonable
simulation time. This may be a consequence of the typically huge scaling
corrections of BD \cite{balfdaar}. However, we believe that the main
reason for those difficulties is the parabolic $\lambda-p$ relation, which
significantly reduces the range of $p$ where both regimes can be numerically
analyzed.

\section{Conclusion}
\label{secconclusion}

We studied a competitive growth model in $1+1$ dimensions involving 
RSOS deposition, with probability $p$, and RSOS evaporation, with probability
$1-p$. This model may be viewed as a discrete realization of the continuum
KPZ equation with an adjustable nonlinear coupling $\lambda$ related to $p$.
Its corresponding KPZ equation is derived, showing that $|\lambda|$ linearly
increases with $q\equiv p-1/2$, so that the process belongs to the EW 
class for $p=1/2$. This result is confirmed numerically by calculation of
tilt-dependent velocities for several values of $p$. It contrasts to the
parabolic $\lambda-p$ relation obtained for competing models involving a KPZ
and an EW process, which shows that this relation, although being of wide
applicability, is not universal.

We also calculated numerically the scaling amplitudes of the EW and KPZ growth
regimes for several values of $p$. From these quantities, estimates of the
crossover
times $t_c$ were obtained. They scale as Eq. (\ref{deffi}) with $\phi=4.1\pm
0.1$, in excellent agreement with the theoretically predicted value of the
crossover exponent. This result improves previous ones, which suggested
$\phi\approx 3$ from simulations of the KPZ equation. We believe that this work
provides an important, possibly definite confirmation of scaling relations
predicted for the EW-KPZ crossover in $d=1$.

\acknowledgments

TJO acknowledges support from CAPES and CNPq and
FDAAR acknowledges support from CNPq and FAPERJ (Brazilian agencies).


\vskip1cm
 
\begin{figure}[!h]
\includegraphics[clip,width=0.9\textwidth, 
height=0.9\textheight,angle=0]{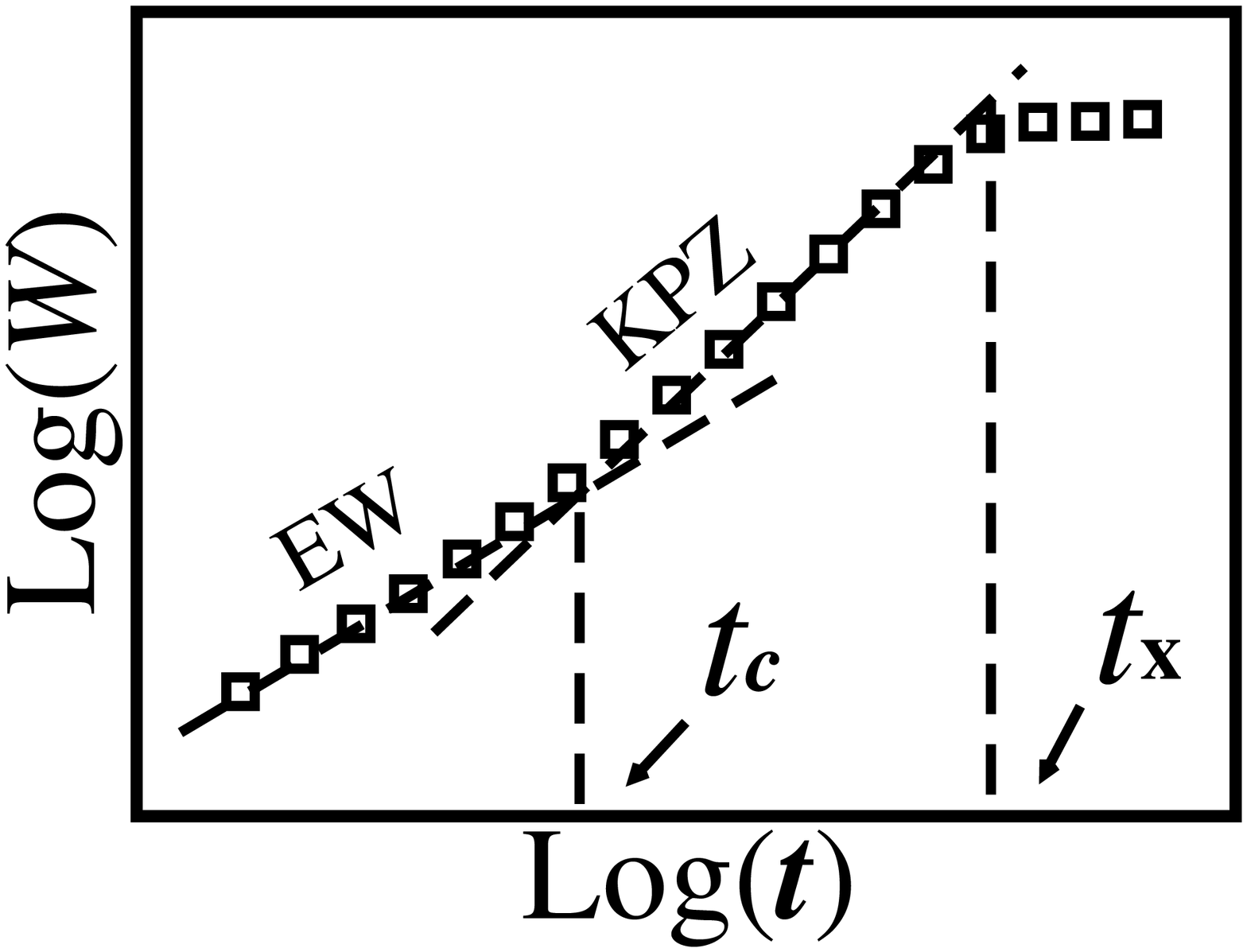}   
\caption{\label{fig1} Typical time evolution of the roughness of a system with
an EW-KPZ crossover at time $t_c$ and saturation time $t_\times$.
}
\end{figure}

\begin{figure}[!h]
\includegraphics[clip,width=0.8\textwidth, 
height=0.8\textheight,angle=0]{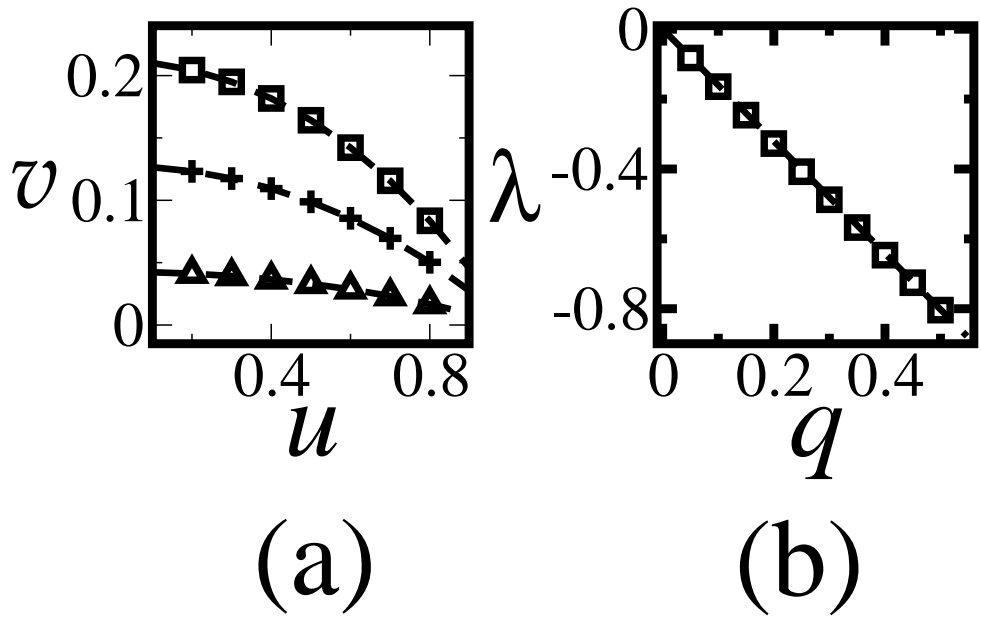}   
\caption{\label{fig2} (a) Growth velocity $v$ as a function of inclination $u$
for the
competitive model with $p=0.55$ (triangles), $p=0.65$ (crosses) and $p=0.75$
(squares). Dashed lines are parabolic fits of each set of data. (b) Estimates of
$\lambda$ as a function of the reduced probability $q\equiv p-1/2$ (squares) and
a least squares fit of the data (dashed line).}
\end{figure}

\begin{figure}[!h]
\includegraphics[clip,width=0.9\textwidth, 
height=0.9\textheight,angle=0]{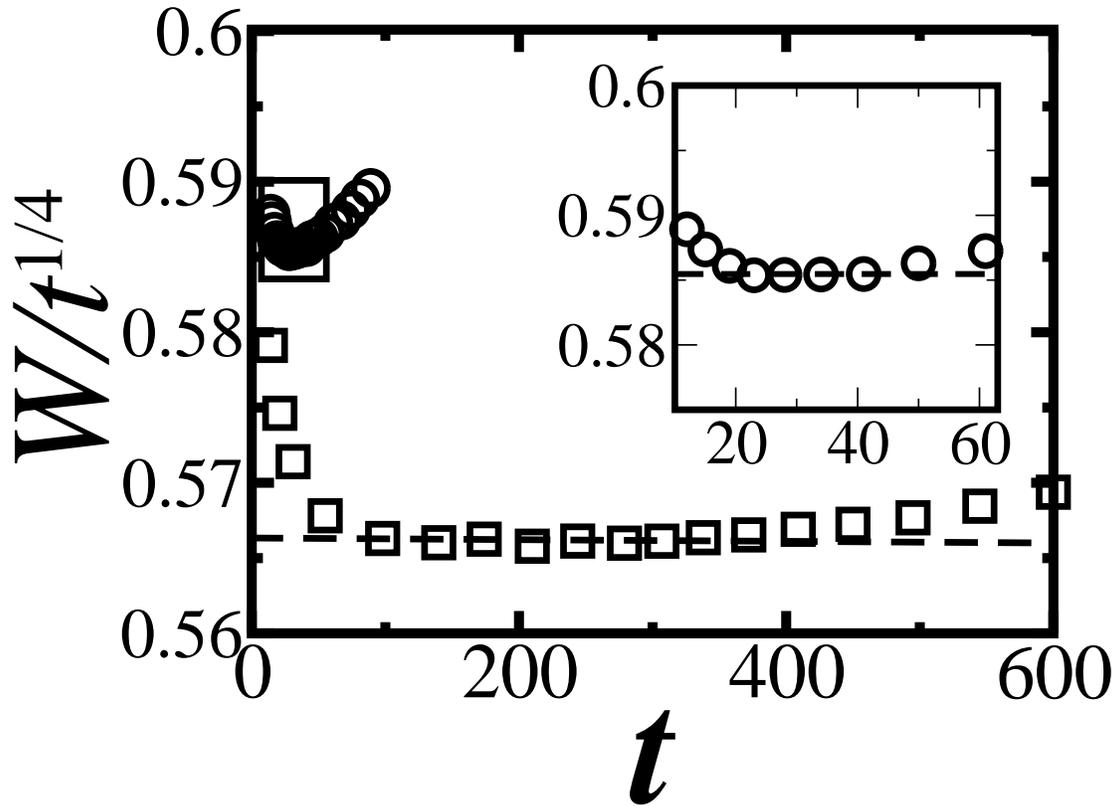}   
\caption{\label{fig3} $W/t^{1/4}$ at small times $t$ for
the competitive model with $p=0.7$ (circles) and $p=0.6$ (squares) in a large
lattice ($L={10}^5$). The inset shows a zoom of the data for $p=0.7$ in the EW
region. The dashed lines are linear fits of the data in those regions.
}
\end{figure}  
 
\begin{figure}[!h]
\includegraphics[clip,width=0.9\textwidth, 
height=0.9\textheight,angle=0]{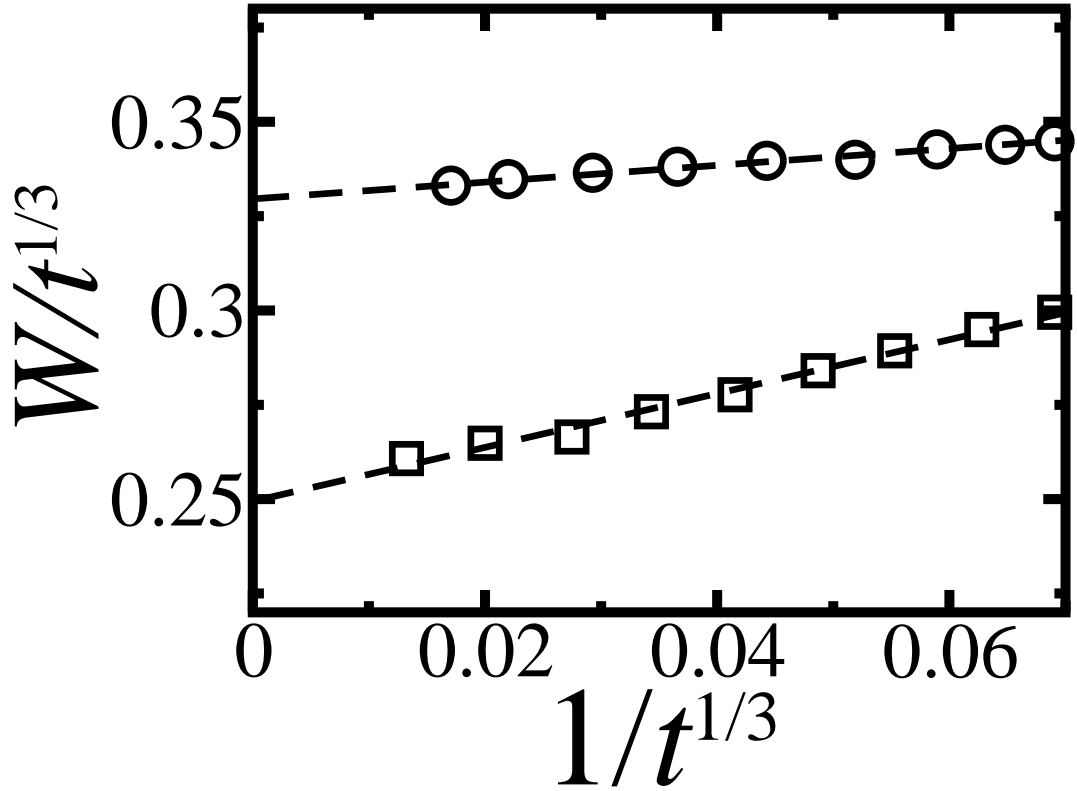}   
\caption{\label{fig4} $W/t^{1/3}$ versus $1/t^{1/3}$ at long times, for $p=0.7$
(circles) and
$p=0.6$ (squares) in a large lattice ($L={10}^5$). The dashed lines are linear
fits of the data.
}
\end{figure}  
 
\begin{figure}[!h]
\includegraphics[clip,width=0.9\textwidth, 
height=0.9\textheight,angle=0]{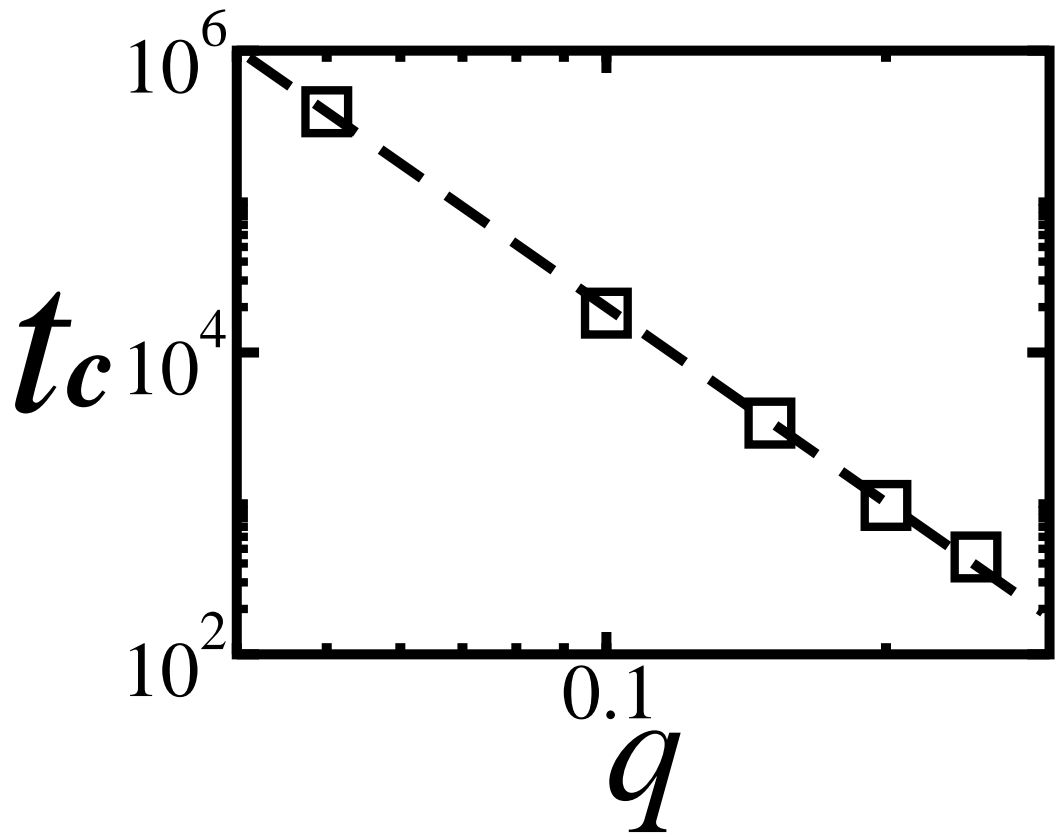}   
\caption{\label{fig5} Crossover time $t_c$ versus $q\equiv p-1/2$ for the
competitive model
with $0.55\leq p\leq 0.8$. The dashed line is a linear fit of the plot, with
slope near $-4.1$.}
\end{figure}       
                        
\end{document}